\documentclass[journal ]{new-aiaa}
\usepackage[utf8]{inputenc}
\usepackage{textcomp}

\usepackage{graphicx}
\usepackage{amsmath,amsfonts}
\newcommand{\NN}{\mathbb{N}}
\newcommand{\RR}{\mathbb{R}}
\newcommand{\LL}{\mathcal{L}}
\usepackage{cleveref}
\usepackage[version=4]{mhchem}
\usepackage{siunitx}
\usepackage{longtable,tabularx}
\usepackage{algorithm}
\usepackage{algorithmicx}
\usepackage{tikz}
\usepackage{circuitikz}
\usetikzlibrary{shapes.geometric}
\usepackage[noend]{algpseudocode}
\title{Online Convex Optimization for On-Board Routing in High-Throughput Satellites}

\author{Olivier Bélanger\footnote{M.A.Sc., Department of Electrical Engineering, \texttt{olivier.belanger@polymtl.ca}}}
\affil{Polytechnique Montréal, Montréal, QC, H3T 0A3, Canada}
\author{Jean-Luc Lupien\footnote{PhD Candidate, Department of Civil and Environmental Engineering, \texttt{jllupien@berkeley.edu}}}
\affil{University of California, Berkley, CA, 94720, United States}
\author{Olfa Ben Yahia\footnote{Dr., Department of Electrical Engineering, \texttt{olfa.ben-yahia@polymtl.ca}}}
\affil{Polytechnique Montréal, Montréal, QC, H3T 0A3, Canada}
\author{Stéphane Martel\footnote{Product Manager, Satellite Systems, \texttt{stephane.martel@mda.space}}}
\affil{MDA Space, Montréal, QC, H9X 3R2, Canada}
\author{Antoine Lesage-Landry\footnote{Associate Professor, Department of Electrical Engineering, Poly-Grames, GERAD \& Mila, \texttt{antoine.lesage-landry@polymtl.ca}}, and Gunes Karabulut Kurt\footnote{Full Professor, Department of Electrical Engineering \& Poly-Grames, \texttt{gunes.kurt@polymtl.ca}}}
\affil{Polytechnique Montréal, Montréal, QC, H3T 0A3, Canada}

\begin{document}

\maketitle

\begin{abstract}
The rise in low Earth orbit (LEO) satellite Internet services has led to increasing demand, often exceeding available data rates and compromising the quality of service. While deploying more satellites offers a short-term fix, designing higher-performance satellites with enhanced transmission capabilities provides a more sustainable solution. Achieving the necessary high capacity requires interconnecting multiple modem banks within a satellite payload. However, there is a notable gap in research on internal packet routing within extremely high-throughput satellites. To address this, we propose a real-time optimal flow allocation and priority queue scheduling method using online convex optimization-based model predictive control. We model the problem as a multi-commodity flow instance and employ an online interior-point method to solve the routing and scheduling optimization iteratively. This approach minimizes packet loss and supports real-time rerouting with a low computational overhead. Our method is tested in numerical simulation on a next-generation extremely high-throughput satellite model, demonstrating its effectiveness compared to a reference batch optimization and to traditional methods.
\end{abstract}

\newpage
\section{Introduction}
In recent years, there has been a significant increase in the utilization of low Earth orbit (LEO) satellite Internet connectivity services. In certain regions, the growing demand outpaces the available data rates, thereby compromising future quality of service (QoS). In fact, some users of the most popular satellite constellation have experienced speed drops of up to 54\% year-over-year~\cite{brodkin2022starlink}. A recent review highlights that this decline in speeds is likely due to growing subscriptions and increased network congestion~\cite{anders2024starlink}. While deploying more satellites is an acceptable short-term solution, a more sustainable alternative involves the design of higher-performance satellites with increased transmission capabilities. These regenerative next-generation payloads will require a very high capacity, achievable through interconnecting multiple modem banks to significantly scale throughput compared to traditional single-processor satellites~\cite{10648786}. Software-based, multi-modem banks architectures, as presented in~\cite{10648786, yahia2024}, aim to increase total throughput and improve efficiency in flow allocation. This approach is particularly relevant for high-throughput satellites (HTSs), very high-throughput satellites (VHTSs), and recently introduced extremely high-throughput satellites (EHTSs)~\cite{10648786}. The multi-processor architecture is essential for providing the computing power required to achieve the promised throughputs of non-geostationary Earth orbit (non-GEO) HTSs. Given the more complex architecture, efficient internal packet routing also becomes central.

Despite the critical need, there is a lack of current research on the internal packet routing between on-board processors in non-GEO HTSs, i.e., the intra-satellite packet routing. Efficient packet routing, in this context, refers to an approach tailored specifically to the unique constraints and requirements of on-board processors within non-GEO HTSs, aiming to minimize information loss and ensure QoS~\cite{electronics8060683}. To the authors' best knowledge, the only existing methods in the literature addressing efficient internal packet routing as defined here are ours~\cite{bélanger2024quality, yahia2024}. Both works provide an optimization-based framework to tackle the challenge. Reference~\cite{yahia2024} proposes a simple multi-commodity flow to minimize the maximum residual capacity of the inter-modem links. Reference~\cite{bélanger2024quality} uses a model predictive control (MPC) approach to optimize the internal routing and scheduling of packets within HTSs and leverages iterative feedback between the optimization process and system state observations.

In parallel, we observe that the intra-satellite routing problem shares structural similarities with traditional packet scheduling and routing in networks on chips (NoCs), where dynamic traffic, congestion, and tight resource constraints must also be managed. Recent work on intelligent NoCs uses reinforcement learning (RL) methods to adapt routing and power policies with the goal of improving throughput and energy efficiency~\cite{benmoussa2022intelligentnocsurvey}. For example, the authors of~\cite{alsahli2021qroutingregionaware} use a region-aware Q-routing framework to learn congestion-aware paths and reduce latency in mesh NoCs. Reference~\cite{fan2022cafeen} employs a cooperative multi-agent RL scheme to coordinate routing and power gating to ensure energy efficiency. These approaches demonstrate the value of adaptive, learning-based routing in dynamic environments, but typically lack formal performance guarantees on optimality and constraint satisfaction, features that are central to our proposed approach.

A significant challenge in implementing an MPC approach -- or any optimization-based approach -- in a non-GEO HTS as in~\cite{bélanger2024quality} is due to limited computational resources~\cite{Schwenzer2021}. Satellite systems are resource-constrained, limited by their low power capabilities~\cite{Ivanov}. Addressing these computational limitations is crucial for the practical deployment of MPC-based solutions.

In this work, we propose an online convex optimization (OCO)-based MPC framework for routing and scheduling packets within a non-GEO HTS, referred to as online convex model predictive control ($\texttt{OCMPC}$). As suggested in~\cite{5153127}, we use an online optimization framework to accelerate our MPC approach. Specifically, we employ the online convex interior-point method from~\cite{lupien2024online} to replace the previously used off-the-shelf optimization solver. OCO is chosen for its low computational load and real-time adaptability, making it well-suited for on-board processing in resource-constrained satellite environments subject to exogenous uncertainty. It allows fast decision-making which in turn enables a better response to this uncertainty.
Previous work, such as~\cite{Jerez2014}, has explored the use of OCO with MPC in embedded systems, focusing on first-order methods and achieving high-speed performance for linear-quadratic MPC problems. Other works have focused on OCO in an MPC-like setting with switching costs and predictions, also developing first-order methods. These approaches utilize the concept of a receding horizon and assume full information on the rolling horizon~\cite{9269418}, increasingly inexact predictions as time progresses~\cite{li2020leveraging}, or stochastic prediction errors~\cite{chen2015onlineconvexoptimizationusing}. Our approach differs by incorporating a second-order method, providing convergence guarantees and more accurate solutions at each time step while only making mild assumptions on future rounds, viz., knowledge of the expected stochastic process. 

Non-GEO HTSs increase the need for these algorithmic considerations because their routing conditions evolve more rapidly than those of traditional GEO satellites. Their close proximity to Earth causes fast orbital motion, leading to strongly time-varying routing constraints and quickly evolving incoming flows. As a result, time-consuming or static optimization strategies become insufficient. Routing methods must instead adapt on short timescales while still respecting the tight computational limits imposed by on-board processing. This context makes such an online, computationally efficient optimization framework particularly suitable for non-GEO HTS on-board routing.

Our specific contributions are as follows:
\begin{itemize}[leftmargin=*]
    \item We introduce the $\texttt{OCMPC}$ framework, leveraging second-order methods and accommodating time-varying constraints to enhance decision-making under uncertainty.
    \item We apply the $\texttt{OCMPC}$ framework in the non-GEO HTS context, marking the first use of OCO for real-time internal routing.
    \item We illustrate the performance of our method in an extensive numerical simulation environment where the incoming packet traffic to a non-GEO HTS is modelled by a Markov-modulated Poisson process (MMPP). Our approach proves to be close to optimal 
    % while remaining practical for real-world deployment in satellite systems.
    while remaining practical for satellite systems.
    % while remaining practical.
\end{itemize}

This work builds upon the foundations in~\cite{bélanger2024quality, yahia2024} by introducing an enhanced algorithm that is implementable and relies on fewer, more realistic assumptions. While we retain the multi-modem architecture established in~\cite{yahia2024}, our approach introduces a more advanced, OCO-based MPC framework, which represents a significant step toward real-world implementation. In addition, unlike~\cite{bélanger2024quality}, which employs a standard MPC framework, we use an online optimization framework to accelerate the solution process at a lower computational cost, paired with a more accurate, MMPP-based model for incoming packet flows. Lastly, we extend the $\varepsilon\texttt{OIPM-TEC}$ algorithm presented by~\cite{lupien2024online} with the notion of MPC, and apply our resulting algorithm to packet routing in non-GEO HTSs.
%rather than the power grid.}

This article is organized as follows: Section \ref{sec:prob_statement} discusses the MPC problem and presents MMPPs. Section \ref{sec:oco_mpc} introduces OCO and details our proposed methodology for integrating these concepts as $\texttt{OCMPC}$. Section \ref{sec:num_res} presents the numerical settings and results and Section \ref{sec:concl} provides closing remarks and outlines future work.

%*********************************%
\section{Problem Statement}\label{sec:prob_statement}
%*********************************%

In~\cite{bélanger2024quality}, we introduced an MPC framework for packet routing and scheduling in HTSs, which demonstrated excellent performance, albeit at a high computational cost. In this work, we address this challenge by developing a more efficient approach, filling a critical gap in the literature for practical implementation in resource-constrained environments.

\subsection{Optimization Formulation}
We first provide the key notation used at a given discretized time increment $t$. Let a satellite contain $M \in \NN$ modem banks and manage $P \in \NN$ priority levels. Each commodity-module pair $(p \in \{1,2,...,P\}, m \in \{1,2,...,M\})$ is associated with a queue $Q_p^m(t) \in \RR^+$ managing packet storage and transmission. Let $w_p^m(t) \in [0,1]$ be the scheduler weight which determines the portion of packets of priority $p$ selected from each queue in modem bank $m$ for processing and transmission between time steps $t$ and $t+1$. Let the net inflow and outflow of commodity $p$ in module $m$ be denoted as $f_p^{\text{in}, m}(t) \in \RR^+$ and $f_p^{\text{out},m}(t)\in \RR^+$, respectively. The queue's inflow and outflow balance, $\Delta Q_p^m(t) \in \RR$, is defined alongside the packet loss ${\mathcal{L}}^m_p(t) \in \RR^+$, the expected incoming demand $\hat{F_p}(t) \in \RR^+$, and the cost incurred by losing a packet of priority $p$, $k_p \in \RR^+$. 
We aim to minimize the total cost-weighted packet loss on each priority $p$ over the time steps in a given MPC with a rolling horizon of length $W \in \NN$, yielding Problem (\ref{eq:mpc_aeroconf}).
\begin{subequations} \label{eq:mpc_aeroconf}
    \begin{align}
    \min_{\substack{w_p^m(\tau), f_p^{\text{in},m}(\tau), \\ \tau \in \{ t,t+1,...,t+W \}}}& \sum_{\tau=t}^{t+W}\sum_{p=1}^{P}\sum_{m=1}^{M} {\mathcal{L}}^m_p(\tau) k_p \label{eq:eq_obj}\\
    % \end{align}
    % \begin{align}
        \text{subject to} \quad \;\;&  f_p^{\text{in}, m}(t) - f_p^{\text{out},m}(t) - \Delta Q_p^m(t) - {\mathcal{L}}^m_p(t) = 0, \label{eq:eq_cstr_1} \\
        & \sum_{p =1}^P w_{p}^m(t) = 1,  \label{eq:eq_cstr_2}\\
        & \sum_{m =1}^M f_p^{\text{in},m}(t) =  \hat{F_p}(t)  ,\label{eq:eq_cstr_3}\\
        & Q_p^m(t+1) = Q_p^m(t) + \Delta Q_p^m(t) , \label{eq:eq_cstr_4}\\
        & Q_p^m(0) = Q_p^m(T-1) = Q_0,  \label{eq:eq_cstr_5}\\
        & 0 \leq w_{p}^m(t) \leq 1,  \label{eq:ineq_cstr_1}\\
        & \big|w_{p}^m(t) - w_{p}^m(t-1)\big| \leq \overline{\Delta w},  \label{eq:ineq_cstr_2}\\
        & f_p^{\text{out},m}(t) \leq \frac{w_p^m(t)}{\Delta s},  \label{eq:ineq_cstr_3} \\
        & \sum_{p =1}^P Q_p^m(t) \leq \overline{Q}^m,  \label{eq:ineq_cstr_4} \\
        & \sum_{p =1}^P f_p^{\text{out},m}(t)  \leq \overline{C}^m, \label{eq:ineq_cstr_5}
    \end{align} 
\end{subequations}
where \eqref{eq:eq_obj} is the convex packet loss cost minimization objective function, \eqref{eq:eq_cstr_1} is the packet balance equation, \eqref{eq:eq_cstr_2} ensures the scheduler weights are normalized, \eqref{eq:eq_cstr_3} matches the routed incoming flow to the observed demand per priority, \eqref{eq:eq_cstr_4} updates the queue occupancy across time, \eqref{eq:eq_cstr_5} sets the initial and final queue occupancy, \eqref{eq:ineq_cstr_1} bounds the scheduler weights, \eqref{eq:ineq_cstr_2} imposes ramp constraints on the scheduler weights, \eqref{eq:ineq_cstr_3} denormalizes the scheduler weights to associate an actual number of packets processed based on the non-GEO HTS parameter $\Delta s > 0$, which acts as the scheduler clock translating the fraction of packets set by $w^m_p(t)$ to a number of processed packets for a time step $t$, \eqref{eq:ineq_cstr_4} limits the total queue occupancy, and \eqref{eq:ineq_cstr_5} restricts the outflow by the transmission bandwidth. 
Detailed explanations about the constraints are provided in~\cite{bélanger2024quality}.

Problem \eqref{eq:mpc_aeroconf} entails solving a multi-period optimization problem to optimality with $P\times M\times W$ constraints at each decision round. This amounts to an important computational load relative to the on-board available resources and may lead to degraded QoS, which is unacceptable for applications requiring high levels of performance and user satisfaction. To address these computational challenges, we develop a more efficient framework that optimizes resource allocation while maintaining high performance. Before presenting this framework, we first introduce MMPPs.

\subsection{Markov-modulated Poisson processes}

We model incoming packet traffic using an MMPP. In~\cite{dellipriscoli2009demand}, MMPPs were used for Internet Protocol traffic prediction in satellite networks, demonstrating their efficiency in capturing the bursty nature of satellite traffic. MMPPs are particularly adapted for processes with irregular bursts of activity combined with predictable patterns, aligning well with the characteristics of satellite traffic~\cite{scott2003}.

An MMPP is defined by a Poisson process with a rate parameter $\lambda(t)$ governed by a Markov chain. The state of the Markov chain at any time $t$ determines the rate $\lambda(t)$. This allows the model to switch between different traffic intensities based on state probabilities.

To the best of our knowledge, MMPPs have not yet been applied to non-GEO HTSs in the context of Internet connectivity. Given the similar traffic patterns observed in terrestrial and other satellite networks, we extend the application of MMPPs to non-GEO HTSs. The inherent ability of MMPPs to capture both predictable periodic patterns and unpredictable bursts in traffic makes them suitable for modelling non-GEO HTS traffic. By incorporating MMPPs, we can achieve more accurate traffic predictions, enhancing the reliability of our model and bringing it a step closer to real-world implementation.
% \vspace{-2mm}

%*********************************%
\section{Online Convex Model Predictive Control }\label{sec:oco_mpc}
%*********************************%

This section provides background on OCO and presents our \texttt{OCMPC} algorithm.

\subsection{Online convex optimization}
To alleviate the computational burden of MPC, we use OCO~\cite{Hazan2022}. OCO is a framework for providing a sequence of decisions in a dynamic environment where the problem changes over time, and is fully observed only after one commits to a decision~\cite{oco_shalev-schwartz}.

In the context of satellite systems, OCO is particularly advantageous because it enables real-time adjustments to routing and scheduling decisions based on current network conditions, in addition to reducing the computational overhead. Our approach does not require solving (\ref{eq:mpc_aeroconf}) to optimality at each time step; instead, we adopt a strategy where only a single infeasible start Newton step is taken, balancing computational efficiency with the quality of the decision as established by the OCO algorithm performance guarantees. This sacrifices a small amount of optimality for a large gain in calculation speed, which is crucial in resource-limited satellites.

We utilize the epsilon-online interior-point method for time-varying equality-constrained ($\varepsilon\texttt{OIPM-TEC}$) optimization~\cite{lupien2024online} for efficient on-board satellite routing. $\varepsilon\texttt{OIPM-TEC}$ guarantees time-averaged optimal decisions on inequality-constrained convex problems with time-varying equality constraints as the time horizon increases, making it well-suited for our problem. $\varepsilon\texttt{OIPM-TEC}$ is a more streamlined version of $\texttt{OIPM-TEC}$~\cite{lupien2024online}, providing an even lighter framework while still offering performance guarantees within an $\varepsilon$-tolerance of the round optima.

As presented in~\cite{lupien2024online}, the optimization problem is formulated in a matrix form. To align more closely with the constraints in (\ref{eq:eq_cstr_1}) $-$ (\ref{eq:ineq_cstr_5}), we express (\ref{eq:mpc_aeroconf}) as (\ref{eq:mpc_mat}), where $\mathbf{x}_t \in \RR^{n\times P\times M\times W}$ contains the $n = 6$ vectorized variables $f_p^{\text{in},m}(t)$, $w_p^m(t)$, ${\mathcal{L}}^m_p(t)$, $f_p^{\text{out},m}(t)$, $Q_p^m(t)$, and $\Delta Q_p^m(t)$ for each pair $(p, m)$ and each time step in $\{t, t+1, ..., t+W\}$:
\begin{align} 
\min_{\mathbf{x}_t} \quad 
& \mathbf{c}^\top\mathbf{x}_t\nonumber\\ 
\text{subject to} \quad & \mathbf{Ax}_t - \mathbf{b}_t = \mathbf{0} \label{eq:mpc_mat}\\ 
& \mathbf{Cx}_t - \mathbf{d} \mkern7mu \leq \mathbf{0}.\nonumber 
\end{align}

The parameters $\mathbf{A}$ and $\mathbf{b}_t$, respectively, represent the multipliers and coefficients of equality constraints (\ref{eq:eq_cstr_1}) $-$ (\ref{eq:eq_cstr_5}), while $\mathbf{C}$ and $\mathbf{d}$, respectively, represent the multipliers and coefficients of the inequality constraints (\ref{eq:ineq_cstr_1}) $-$ (\ref{eq:ineq_cstr_5}). This formulation is directly applicable to our non-GEO HTS on-board routing problem, where the time-varying constraints, particularly \eqref{eq:eq_cstr_3} and \eqref{eq:eq_cstr_4}, are expressed as equality constraints. Problem~(\ref{eq:mpc_mat}) is obtained by stacking the MPC decision variables across the rolling horizon into a single vector $\mathbf{x}_t$, which includes the flows, scheduler weights, and queue states, for each priority, modem, and time step. The detailed derivation of (\ref{eq:mpc_mat}) from (\ref{eq:mpc_aeroconf}) is presented in the Appendix.

$\varepsilon\texttt{OIPM-TEC}$ solves a problem of the form (\ref{eq:mpc_mat}) by taking an infeasible start Newton step towards optimality and directly observing the impact of that decision. It ensures feasibility by maintaining strict feasibility for time-invariant inequality constraints and sublinearly bounding the violation of time-varying equality constraints under some assumptions. The high-speed movement of LEO satellites introduces dynamism primarily through variations in the incoming flow $F(t)$, affecting $\mathbf{b}_t$. By integrating $\varepsilon\texttt{OIPM-TEC}$ into our MPC framework, we effectively handle these variations while reducing computational overhead and adapting to changing conditions in real time.

The process and interactions defining our approach for HTS internal routing are illustrated in Figure \ref{fig:ocmpc_framework}.

\begin{figure}[H]
    \centering
    \resizebox{1\textwidth}{!}{
    \begin{circuitikz}
        \tikzstyle{every node}=[font=\footnotesize]
        
        % Expected Flows
        \node [font=\footnotesize] (expected) at (0,4) {Expected flow $\hat{F}(t)$};
        
        % Optimization
        \node [draw, rectangle, minimum width=4.5cm, minimum height=1.4cm] (optimization) at (5,4) {
            \begin{tabular}{c}
                Online Optimization over window $W$ \\
                {\footnotesize (via $\varepsilon\texttt{OIPM-TEC}$, 1 Newton Step)}
            \end{tabular}
        };
        \draw [->, >=Stealth] (expected) -- (optimization);
        
        % Constraints
        \node [draw, diamond, aspect=2] (constraints) at (10,4) {Constraints};
        \draw [->, >=Stealth] (constraints) -- (optimization);
        
        % System Parameters
        \node [draw, rectangle, minimum width=3cm, minimum height=1cm] (parameters) at (10,6) {System Parameters};
        \draw [->, >=Stealth] (parameters) -- (constraints);
        
        % Optimized Weights
        \node [draw, rectangle, minimum width=3cm, minimum height=1.0cm] (weights) at (5,1.5) {$f_p^{\text{in},m}(t)$, $w_p^m(t)$};
        \draw [->, >=Stealth] (optimization) -- (weights);
        
        % Satellite System
        \node [draw, rectangle, minimum width=3cm, minimum height=1cm] (satellite) at (10,1.5) {Satellite System};
        \draw [->, >=Stealth] (weights) -- (satellite);
        
        % Realized Flows
        \node [font=\footnotesize] (realized) at (10,-0.5) {Realized flow $F(t)$};
        \draw [->, >=Stealth] (realized) -- (satellite);
        
        % Feedback
        \draw [->, >=Stealth] (satellite) -- (constraints);
        \node at (10.15,2.75) [right] {Feedback};
        
    \end{circuitikz}
    }%
    \caption{$\texttt{OCMPC}$ framework for HTS internal routing}
    \label{fig:ocmpc_framework}
\end{figure}

% \vspace{-2mm}
\subsection{Algorithm}
In this section, we present our methodology in detail. We focus on the algorithmic integration of OCO with MPC for satellite on-board routing. 

Let $\phi:\RR^{n\times P\times M\times W} \to \RR$ be the log-barrier functional as introduced in~\cite{boyd2004convex} and let $\eta > 0$ be the barrier parameter. As presented in~\cite{boyd2004convex}, interior-point methods solve a problem in the form of 
\eqref{eq:obj_barrier} instead of directly solving one similar to~(\ref{eq:mpc_mat}):
\begin{equation}\label{eq:obj_barrier}
    \min_{\mathbf{x}_t} \quad d_\eta(\mathbf{x}_t),
\end{equation} 
where \( d_\eta(\mathbf{x}_t) =  \eta\mathbf{c}^\top\mathbf{x}_t + \phi(\mathbf{x}_t) \) is the log-barrier functional-augmented objective function ~\cite{lupien2024online}. 
From~\cite{bélanger2024quality},~\cite{lupien2024online},~\cite[equations 10.21 and 10.22]{boyd2004convex}, we define $\texttt{OCMPC}$ in Algorithm \ref{alg:oco_mpc}, where we combine $\varepsilon\texttt{OIPM-TEC}$ and our MPC framework.

While $\varepsilon\texttt{OIPM-TEC}$ is effective in many scenarios, it is only guaranteed to respect equality constraints under conditions more stringent than can be assumed for this application. To mitigate this potential issue, we introduce a feedback correction mechanism that proportionally adjusts the flow allocation \( f_p^{\text{in}, m}(t) \) for all pairs $(p, m)$ so as to enforce the incoming-flow demand specified in constraint~\eqref{eq:eq_cstr_3}. The mechanism consists of proportionally rescaling the tentative decision, ensuring feasibility at negligible computational cost while preserving the structure of the solution returned by \texttt{OCMPC}. This correction is applied once the realized incoming flow is observed, compensating for the fact that the decision $\mathbf{x}_t$ is computed in a predictive manner, prior to the realization of the actual packet arrivals.

\begin{algorithm}[tb]
\caption{$\texttt{OCMPC}$ for online on-board routing}\label{alg:oco_mpc}
\begin{algorithmic}[1]
    \State \textbf{Parameters:} $T$, $W$, $\mathbf{A}$.
    \State \textbf{Initialization:} Given $\mathbf{x}_0$ and $\eta$.
    % \State Define $\tilde{\mathbf{x}}_0 = \mathbf{x}_0$
    \For{$t$ in $\{0,1, \dots,T-1\}$}
        \State Observe $\hat{F_p}(t)$.
        \State Implement the decision $w_p^m(t)$ and $f_p^{\text{in}, m}(t)$ from $\mathbf{x}_t$.
        \If{$\mathbf{x}_t$ is not such that $\sum_{m =1}^M f_p^{\text{in},m}(t) = F_p(t)$ 
        \Statex \hspace{1.15em} $ \forall \; p \in \{1, 2, \dots, P\}$}
        \State Apply the feedback correction.
        \EndIf
        \State Observe the outcome $\mathbf{c}^\top\mathbf{x}_t$, the new constraint $\mathbf{b}_t$, and \Statex \hspace{1.15em} the new states $Q_p^m(t)$, $\Delta Q_p^m(t)$, ${\mathcal{L}}^m_p(t)$, and $f_p^{\text{out},m}(t)$.
        \State \underline{Update decision}: 
        \State ${\begin{bmatrix} \mathbf{x}_{t+1} \\ - \end{bmatrix}}  = {\begin{bmatrix} \mathbf{x}_t \\ - \end{bmatrix}} - {\begin{bmatrix} \nabla^2\phi(\mathbf{x}_t) & \mathbf{A}^\top \\ \mathbf{A} & \mathbf{0} \end{bmatrix}}^{-1} {\begin{bmatrix} \nabla d_{\eta}(\mathbf{x}_t) \\ \mathbf{Ax}_t - \mathbf{b}_t \end{bmatrix}}.$
    \EndFor
\end{algorithmic}
\end{algorithm}

The sequence $\{\mathbf{x}_t\}^T_{t=1}$ provided by Algorithm~\ref{alg:oco_mpc} has provable bounds on dynamic $\varepsilon$-regret and constraint violations under certain conditions. Interested readers are referred to~\cite{lupien2024online} for further details.

Finally, we note that alternative fast solution strategies could be considered. First-order online optimization methods~\cite{zinkevich2003online}, for instance, typically offer lower per-iteration computational cost, but often require projection steps onto the feasible set and exhibit weaker performance guarantees compared to second-order approaches. Learning-based methods, e.g. RL~\cite{benmoussa2022intelligentnocsurvey,alsahli2021qroutingregionaware,fan2022cafeen}, represent another possible direction, as they can shift part of the computational burden to an offline training phase; however, they generally do not provide provable guarantees on performance or constraint satisfaction. For these reasons, we focus on a second-order online optimization method that offers a favorable balance between computational efficiency and theoretical guarantees.

%****************************%%
\section{Numerical Results}\label{sec:num_res}
%****************************%%
This section presents the simulation setup and numerical results.

\subsection{Numerical setting}

Let $F(t)$ be the incoming flow, modelled as a discrete-time Markov chain with state space $\mathcal{S}=\{1,2,3\}$, representing three different traffic states. Consider the transition probability matrix $P_\lambda$ defined as:
% \vspace{-1mm}
\[
P_\lambda = \begin{bmatrix}
    0.8 & 0.15 & 0.05 \\ 0.1 & 0.8 & 0.1 \\ 0.05 & 0.2 & 0.75
\end{bmatrix}.
\]
Each state $ i \in \mathcal{S} $ is associated with a Poisson process characterized by a rate parameter $\lambda_i$ packets per time increment, where  $\lambda_1 = 20$, $\lambda_2 = 25$, and $\lambda_3 = 30$. The non-GEO HTS determines $P_\lambda$ based on its relative position to Earth, reflecting the probability of transitioning between different traffic states. At any given time $t$, the traffic intensity follows a Poisson distribution with rate $\lambda_{F(t)}$, modulated by the current state $F(t)$ of the Markov chain.

We assume that a given non-GEO HTS knows its position relative to Earth and can therefore estimate which traffic intensity $\lambda_i$ it will experience at any given time. This means that at each time step $t$, the expected incoming flow $\hat{F}_p(t)$ in \eqref{eq:eq_cstr_3} will be one of the three $\lambda_{F(t)}$ values, i.e., $\hat{F}_p(t) \in \{ \lambda_1, \lambda_2, \lambda_3 \}$. 

This approach allows us to realistically model the temporal fluctuations in satellite Internet traffic. Simulating the traffic as an MMPP lets us capture the inherent stochasticity and time-dependent behaviour of the system. Because we focus on packet routing within a single non-GEO HTS, the configuration parameters in Table~\ref{tab:experimental_setup_OCO_MPC} remain constant, with only $F(t)$ varying temporally.

\begin{table}[tb]
    \centering
    \footnotesize
    \caption{Simulation setup parameters
    of the $\texttt{OCMPC}$ framework}
    \label{tab:experimental_setup_OCO_MPC}
    
    \begin{tabular}{lp{4cm}p{6.5cm}}
    % \begin{tabular}{c|c|c}
        \hline
        \textbf{Parameter} & \textbf{Value} & \textbf{Description} \\
        \hline
        $T$ & 100 time steps & Time horizon of the simulation \\
        $W$ & 5 time steps & MPC window size \\
        $M$ & 16 & Number of modem banks \\
        $P$ & 3 & Number of distinct priorities \\
        $k_p$ & 10, 4, 1 & Packet loss costs for high, medium, and low priority packets \\
        $\overline{Q}$ & 10 packets & Maximum queue size \\
        $Q_0$ & 0 & Initial and final queue occupancy \\
        $\overline{\Delta w}$ & 10\% & Maximum deviation of scheduler weights \\
        $\lambda_p$ & $\{20, 25, 30\}$ packets/time step & Average arrival rate, following an MMPP \\
        $\eta$ & $10^4$ & Barrier parameter for the $\varepsilon \texttt{OIPM-TEC}$ algorithm \\
        \hline
    \end{tabular}
    
\end{table}

The simulation setup spans a time horizon of $T=100$ time steps and uses an MPC window of $W=5$.
% similarly to~\cite{bélanger2024quality}. 
We consider a satellite equipped with $M=16$ modem banks~\cite{yahia2024}, each capable of handling $P=3$ distinct data priorities, e.g., voice over Internet Protocol (high priority,~$p=1$), instant messaging (medium priority,~$p=2$), and emails (low priority,~$p=3$). The packet loss costs are set to 4, 2, and 1 arbitrary units, respectively. We normalize each $\lambda_i$ by the packet loss cost $k_p$ to create an inverse relationship with respect to priorities. Each queue can hold a maximum of $\overline{Q} = 10$ packets, with both the initial and final states of the queues being empty ($Q_0 = 0$) to ensure continuity. The scheduler weights are restricted to change by no more than $\overline{\Delta w}$ = 10\% between consecutive time steps. The entire simulation framework is designed to be scalable, enabling the representation of flow values across various magnitudes, such as $10^9$ packets in the context of a non-GEO HTS. 
The use of $\varepsilon\texttt{OIPM-TEC}$ requires the setting of specific parameters. We initialize the initial guess $\mathbf{x}_0$ randomly and verify its feasibility before applying the Newton step. Additionally, we set the barrier parameter~$\eta$ to~$10^4$, which translates to $\varepsilon \sim \frac{O(N)}{\eta}$, where $N = 6(W+1)MP$ is the dimension of the decision variable~$\mathbf{x}_t$~\cite[Theorem 2]{lupien2024online}. Simulation parameters are summarized in Table \ref{tab:experimental_setup_OCO_MPC}.

We conduct 100 Monte Carlo simulations to ensure consistency and to account for the stochastic nature of the MMPP-based incoming flow. 
We use the following methods to benchmark our $\texttt{OCMPC}$ algorithm: 

\begin{itemize}
    \item \emph{Batch with hindsight}: Problem (\ref{eq:mpc_aeroconf}) solved with hindsight information on the incoming flow~\cite{yahia2024}, serving as the best yet unreachable performance;
    \item \emph{MPC}: Our MPC framework solved to optimality at each round~\cite{bélanger2024quality}, providing a reference for our online optimization algorithm, though being computationally expensive;
    \item \emph{Cost-proportional allocation}: A rule-based controller that sets the scheduler weights proportional to the associated cost \( k_p \).
\end{itemize}

% \vspace{-4mm}
\subsection{Numerical results}

\begin{figure}[tb]
    \centering
    \includegraphics[ width=\columnwidth]{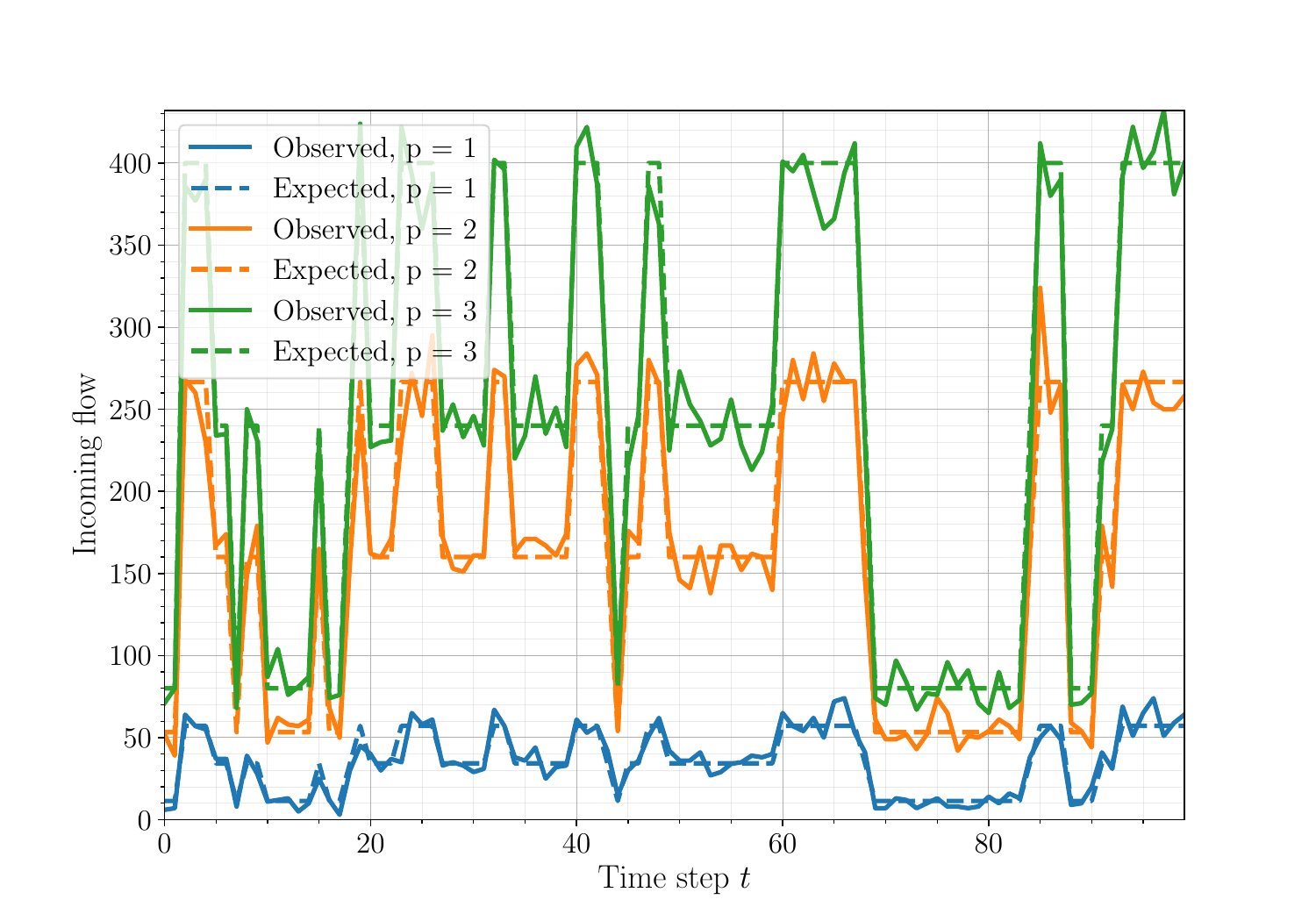}
    \caption{MMPP-based incoming flows (packets) across time for different priorities}
    \label{fig:f_in_vs_t}
\end{figure}

The incoming flow distribution, depicted in Figure \ref{fig:f_in_vs_t}, illustrates the flows across priorities, highlighting the system's inherent uncertainty when modelled with an MMPP. While the MMPP captures the average behaviour of the incoming flows, the realized flows remain difficult to predict at each step and fluctuate around their expected values. This short-term variability, present across priorities and across tests, directly contributes to the variability observed in performance metrics and underscores the need for online methods that can adapt to the realized traffic.
% \vspace{-5mm}
% \vspace{-8mm}
\begin{figure}[tb]
    \centering
    \includegraphics[width=\columnwidth]{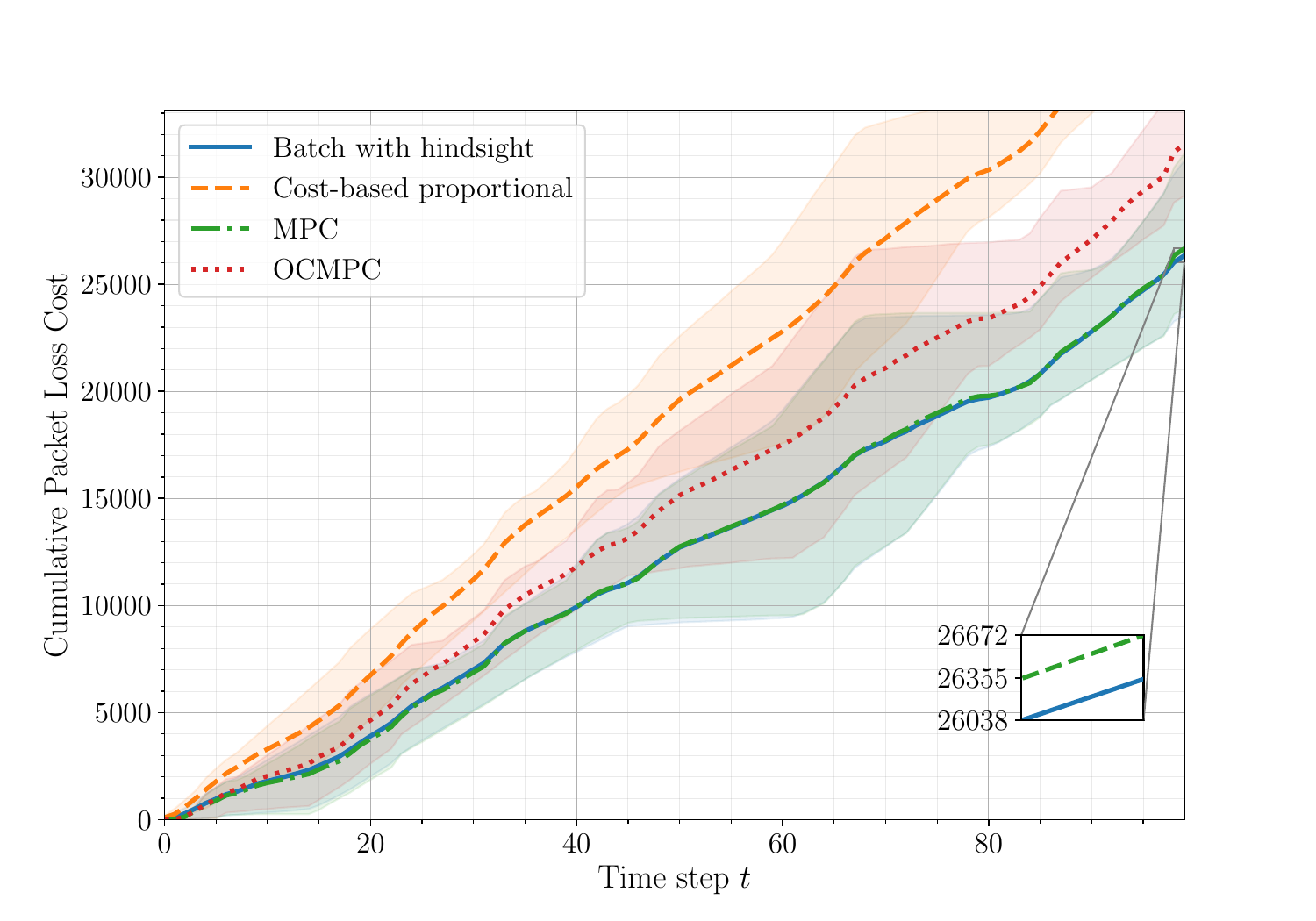}
    \caption{Comparison of cumulative packet loss costs across different routing methods over time averaged over 100 Monte Carlo runs}
    \label{fig:L_cost_vs_t_oco}
\end{figure}

Figure~\ref{fig:L_cost_vs_t_oco} illustrates the performance of our \texttt{OCMPC} algorithm against the other benchmarks. The cumulative packet loss costs are averaged over 100 Monte Carlo runs to ensure statistical significance. The shadowed regions depict the central 95\% interval of the data, highlighting the variability and uncertainty in the system's performance across the Monte Carlo simulations. 

By design, the \textit{batch with hindsight} method achieves the best performance because it uses full information on the incoming flow. The MPC framework performs closely, incurring only 1.24\% more packet loss costs than the \textit{batch with hindsight} method. This underscores the relevance of applying such a framework to flow routing on-board a non-GEO HTS. Our $\texttt{OCMPC}$ approach also performs well, with 19.73\% more packet loss costs than the \textit{batch with hindsight} method and 17.91\% more than the comparative MPC method. While the $\texttt{OCMPC}$ algorithm trades off some performance for efficiency, the $\texttt{OCMPC}$ algorithm reduces computational complexity, making it suitable for real-time applications. It uses a single Newton step per iteration towards the optima and does not rely on a full-on solver like~\cite{mosek}, making it readily implementable while maintaining good performance. This balance between performance and computational efficiency is crucial for satellite systems with limited processing power.
Let $\epsilon > 0$ be the convergence tolerance of the optimization solver. For a linear MPC, the complexity is $O(n^3m|\log(\epsilon)|)$, as a single linear problem solving step costs $O(n^2m)$ per iteration~\cite{Tits2006ConstraintReduction}, and interior-point methods used in CVXPY require, as a worst-case upper bound, $O(n|\log(\epsilon)|)$ operations~\cite{nocedal2006numerical}, where $n$ and $m$ are the number of variables and constraints, respectively. In contrast, $\texttt{OCMPC}$ performs only a single iteration, reducing the complexity to $O(n^2m)$ per time step and, therefore, significantly increasing computational efficiency.  While batch optimization solves a larger problem than MPC, it does so only once, making it less complex but unrealistic, as previously explained. Conversely, the greedy method is computationally efficient ($O(1)$), but ineffective as shown in Figure \ref{fig:L_cost_vs_t_oco}.

The $\texttt{OCMPC}$ approach significantly outperforms the \textit{proportional approach}, which registered the highest costs with a total of 49.27\% more than the hindsight optimum. While the \textit{proportional approach} is computationally lightweight, it performs poorly. The poor performance of the \textit{proportional approach} underscores the importance of strategic algorithm design in achieving effective flow management. This contrast highlights the advantage of utilizing the limited on-board processing capabilities by implementing the $\texttt{OCMPC}$ approach.

In summary, our $\texttt{OCMPC}$ achieves a good compromise: it closely tracks the fully optimal MPC (within $\sim$18\% of its cost) and vastly outperforms the simpler heuristic, validating our design goal. These results indicate that \texttt{OCMPC} provides a competitive level of performance despite relying on a much simpler update rule than full MPC. This efficiency arises from performing only a single second-order update per time step, rather than solving a complete convex program, making the method lightweight enough for real-time applications.

% \vspace{-4mm}
%****************************%%
\section{Conclusion}\label{sec:concl}
%****************************%%

In this article, we propose a novel algorithm to address the challenge of on-board routing in a non-GEO HTS equipped with multiple modem banks. Combining the predictive capabilities of MPC with the implementability of $\varepsilon\texttt{OIPM-TEC}$, our $\texttt{OCMPC}$ approach provides performance comparable to MPC and \emph{batch optimization with hindsight}, while being both implementable and computationally efficient. Taken together, these results illustrate that a second-order online optimization method can serve as a practical, high-performing algorithm for real-time on-board routing in non-GEO HTSs. Because these systems experience rapidly varying incoming traffic caused by their fast movement relative to Earth, the ability of \texttt{OCMPC} to react quickly to these fluctuations is particularly valuable. Finally, this work highlights that such second-order methods can be embedded within next-generation non-GEO HTSs, enabling fast on-board routing decision-making.

Future work could exploit the fast computation provided by $\varepsilon\texttt{OIPM-TEC}$ to further reduce the time step duration, closely tracking incoming flow and enhancing performance, while maintaining computational efficiency. This could be applied to a larger-scale internal satellite topology, such as a toroidal structure proposed in~\cite{yahia2024}, or even a dynamically reconfigurable one. Additionally, the approach could be extended and adapted to MEO or GEO satellites, as well as non-satellite-based telecommunication networks, examining how their specific constraints and flow behaviour influence the algorithm's performance. Another avenue is to examine varying packet lengths, providing insights into their impact on routing efficiency.
A further direction for future work is to conduct a systematic comparison with reinforcement learning-based routing approaches in order to better understand the trade-offs between empirical adaptability and formal performance and constraint guarantees. Lastly,  the evaluation of empirical computation times on representative on-board processors would provide valuable insights. While absolute runtimes are hardware dependent, such tests would complement the hardware-agnostic complexity analysis reported here and further quantify real-time feasibility.

\newpage

\section*{Appendix}
This appendix details the mathematical formulation presented in \eqref{eq:mpc_mat}. Recall Problem (\ref{eq:mpc_aeroconf}), where the equality and inequality constraints are now presented and regrouped in the form of \eqref{eq:mpc_mat}:

\begin{subequations} \label{eq:mpc_aeroconf_mat}
    \begin{align}
    \min_{\substack{w_p^m(\tau), f_p^{\text{in},m}(\tau), \\ \tau \in \{ t,t+1,...,t+W \}}}& \sum_{\tau=t}^{t+W}\sum_{p=1}^{P}\sum_{m=1}^{M} {\mathcal{L}}^m_p(\tau) k_p \label{eq:eq_obj_mat} \\
    % \end{align}
    % \begin{align}
        \text{subject to} \quad \;\;&  f_p^{\text{in}, m}(t) - {\mathcal{L}}^m_p(t) - f_p^{\text{out},m}(t) - \Delta Q_p^m(t) &&= 0, \label{eq:eq_cstr_1_mat} \\
        & \sum_{p =1}^P w_{p}^m(t) - 1 &&= 0,  \label{eq:eq_cstr_2_mat}\\
        & \sum_{m =1}^M f_p^{\text{in}, m}(t) - \hat{F_p}(t) &&= 0  ,\label{eq:eq_cstr_3_mat}\\
        & -Q_p^m(t) + Q_p^m(t+1) - \Delta Q_p^m(t) &&= 0 , \label{eq:eq_cstr_4_mat}\\
        & w_{p}^m(t-1) - \hat{w}_{p}^m(t-1) &&= 0,  \label{eq:eq_cstr_5_mat}\\
        & Q^m_p(t) - Q_{p,\text{sys}}^m(t) &&= 0,  \label{eq:eq_cstr_6_mat}\\
        &-w_{p}^m(t) &&\leq 0,  \label{eq:ineq_cstr_1_1_mat}\\
        & w_{p}^m(t) -1 &&\leq 0,  \label{eq:ineq_cstr_1_2_mat}\\
        & w_{p}^m(t) - w_{p}^m(t-1) - \overline{\Delta w} &&\leq 0, \label{eq:ineq_cstr_2_1_mat}\\
        & -w_{p}^m(t) + w_{p}^m(t-1) - \overline{\Delta w} &&\leq 0, \label{eq:ineq_cstr_2_2_mat}\\
        & -\frac{w_p^m(t)}{\Delta s} + f_p^{\text{out},m}(t) && \leq 0,  \label{eq:ineq_cstr_3_mat} \\
        & \sum_{p =1}^P Q_p^m(t) - \overline{Q}^m &&\leq 0,  \label{eq:ineq_cstr_4_mat} \\
        & \sum_{p =1}^P f_p^{\text{out},m}(t) - \overline{C}^m  &&\leq 0, \label{eq:ineq_cstr_5_mat}
    \end{align} 
\end{subequations}

To create the decision vector $\mathbf{x}_t \in \mathbb{R}^{n \times P \times M \times (W+1)}$ , we introduce a few definitions. We note that it contains all $n = 6$ optimization variable types used to model the internal dynamics of EHTS:
$f_p^{\text{in},m}(t)$, $w_p^m(t)$, $\LL^m_p(t)$, $f_p^{\text{out},m}(t)$, $Q_p^m(t)$, and $\Delta Q_p^m(t)$. Each variable type is grouped in a matrix for a given type step $t$, e.g.:
\begin{equation*}
F_p^{\text{in},m}(t) = 
    \begin{bmatrix}
        f_1^{\text{in},1}(t) & f_2^{\text{in},1}(t) & \cdots & f_P^{\text{in},1}(t)  \\
        f_1^{\text{in},2}(t) & f_2^{\text{in},2}(t) & \cdots & f_P^{\text{in},2}(t)  \\
        \vdots  & \vdots  & \ddots & \vdots  \\
        f_1^{\text{in},M}(t) & f_2^{\text{in},M}(t) & \cdots & f_P^{\text{in},M}(t)  \\
    \end{bmatrix}_{M \;\times\; P}.
\end{equation*}

We define the operator $\mathrm{vec}:\RR^{m \times n} \mapsto \RR^{mn}$ as follows:

\begin{equation*}
\mathrm{vec}\left(F_p^{\text{in},m}(t)\right) = 
    \begin{bmatrix}
        f_1^{\text{in},1}(t) \\
        f_1^{\text{in},2}(t) \\
        f_1^{\text{in},M}(t) \\
        f_2^{\text{in},1}(t) \\
        f_2^{\text{in},2}(t) \\
        \vdots \\
        f_2^{\text{in},M}(t) \\
        \vdots \\
        f_P^{\text{in},M}(t) 
    \end{bmatrix}_{MP \;\times\; 1}.
\end{equation*}

We define the decision vector $\mathbf{x}_t$ for a given time window ranging from time steps $t-1$ to $t + W - 1$ as :
 
\begin{equation*} \label{eq:x}
\mathbf{x}_t = 
    \begin{bmatrix}
        \text{vec}(F_p^{\text{in},m}(t-1)) \\
        \text{vec}(W_p^m(t-1)) \\
        \text{vec}({\mathcal{L}}_p^m(t-1)) \\
        \text{vec}(F_p^{\text{out},m}(t-1)) \\
        \text{vec}(Q_p^m(t-1)) \\
        \text{vec}(\Delta Q_p^m(t-1)) \\
        \text{vec}(F_p^{\text{in},m}(t)) \\
        \text{vec}(W_p^m(t)) \\
        \text{vec}({\mathcal{L}}_p^m(t)) \\
        \text{vec}(F_p^{\text{out},m}(t)) \\
        \text{vec}(Q_p^m(t)) \\
        \text{vec}(\Delta Q_p^m(t)) \\
        \text{vec}(F_p^{\text{in},m}(t+1)) \\
        \text{vec}(W_p^m(t+1)) \\
        \text{vec}({\mathcal{L}}_p^m(t+1)) \\
        \vdots \\
        \text{vec}(F_p^{\text{out},m}(t+W-1)) \\
        \text{vec}(Q_p^m(t+W-1)) \\
        \text{vec}(\Delta Q_p^m(t + W -1)) 
    \end{bmatrix}.
\end{equation*}

Equality constraints \eqref{eq:eq_cstr_1_mat} $-$ \eqref{eq:eq_cstr_6_mat} are then embedded in $\mathbf{A}$ and $\mathbf{b}_t$. Let the number of columns of $\mathbf{A}$ be the number of decision variables $nPM(W+1)$. Let the rows of $\mathbf{A}$ be the number of individual constraints. Constraints \eqref{eq:eq_cstr_1_mat}, \eqref{eq:eq_cstr_4_mat}, and \eqref{eq:eq_cstr_4_mat} lead to a new row for each combination of $m$ and $p$, whereas \eqref{eq:eq_cstr_2_mat} leads to $M$ rows, and \eqref{eq:eq_cstr_3_mat}, to $P$ rows. This holds for each time step in $\{t-1, \dots, t+W-1\}$. The column vector $\mathbf{b}_t$, which has the same number of rows as $\mathbf{A}$, contains the parameters associated to the time-varying equality constraints \eqref{eq:eq_cstr_3_mat}, \eqref{eq:eq_cstr_5_mat} , and \eqref{eq:eq_cstr_6_mat}, as well as those for the time-invariant equality constraints \eqref{eq:eq_cstr_1_mat}, \eqref{eq:eq_cstr_2_mat}, and \eqref{eq:eq_cstr_4_mat}. Similarly, $\mathbf{C}$ and $\mathbf{d}$ together model the inequality constraints \eqref{eq:ineq_cstr_1_1_mat} $-$ \eqref{eq:ineq_cstr_5_mat}. Like $\mathbf{A}$, $\mathbf{C}$ has $nPM(W+1)$ columns. Constraints~\eqref{eq:ineq_cstr_1_1_mat}~$-$~\eqref{eq:ineq_cstr_3_mat}, \eqref{eq:ineq_cstr_5_mat} lead to $PM$ rows per time step, and \eqref{eq:ineq_cstr_4_mat} leads to $M$ rows per time step. Meanwhile, the column vector $\mathbf{d}$, which has the same number of rows as $\mathbf{C}$, contains the parameters associated to the time-invariant inequality constraints~\eqref{eq:ineq_cstr_1_1_mat}$-$\eqref{eq:ineq_cstr_5_mat}. Finally, to ensure a closed, compact feasible space, we numerically introduce box constraints for each decision variable.

Matrices $\mathbf{A}$ and $\mathbf{C}$ and vectors $\mathbf{b}_t$, $\mathbf{d}$, and $\mathbf{c}^\top$ are now presented. For simplicity, all elements of A are represented for $W=2$. We define $\mathbf{A}$ as:
\setcounter{MaxMatrixCols}{20}
\begin{equation*}
\scriptsize
\mathbf{A} =  
    \begin{bmatrix}
        \mathbf{0} & \mathbf{0} & \mathbf{0} & \mathbf{0} & \mathbf{0} & \mathbf{0} & \mathbf{\delta_{mm'}\delta_{pp'}} & \mathbf{0} & \mathbf{-\delta_{mm'}\delta_{pp'}} & \mathbf{-\delta_{mm'}\delta_{pp'}} & \mathbf{0} & \mathbf{-\delta_{mm'}\delta_{pp'}}  & \mathbf{0} & \mathbf{0} & \mathbf{0} & \mathbf{0} & \mathbf{0} & \mathbf{0} \\
        
        \mathbf{0} & \mathbf{0} & \mathbf{0} & \mathbf{0} & \mathbf{0} & \mathbf{0} & \mathbf{0} & \mathbf{v_m} & \mathbf{0} & \mathbf{0} & \mathbf{0} & \mathbf{0} & \mathbf{0} & \mathbf{0} & \mathbf{0} & \mathbf{0} & \mathbf{0} & \mathbf{0} \\
        
        \mathbf{0} & \mathbf{0} & \mathbf{0} & \mathbf{0} & \mathbf{0} & \mathbf{0} & \mathbf{v_p} & \mathbf{0} & \mathbf{0} & \mathbf{0} & \mathbf{0} & \mathbf{0} & \mathbf{0} & \mathbf{0} & \mathbf{0} & \mathbf{0} & \mathbf{0} & \mathbf{0} \\
        
        \mathbf{0} & \mathbf{0} & \mathbf{0} & \mathbf{0} & \mathbf{0} & \mathbf{0} & \mathbf{0} & \mathbf{0} & \mathbf{0} & \mathbf{0} & \mathbf{\delta_{mm'}\delta_{pp'}} & \mathbf{\delta_{mm'}\delta_{pp'}} & \mathbf{0} & \mathbf{0} & \mathbf{0} & \mathbf{0} & \mathbf{-\delta_{mm'}\delta_{pp'}} & \mathbf{0} \\
        
        \mathbf{0} & \mathbf{\delta_{mm'}\delta_{pp'}} & \mathbf{0} & \mathbf{0} & \mathbf{0} & \mathbf{0} & \mathbf{0} & \mathbf{0} & \mathbf{0} & \mathbf{0} & \mathbf{0} & \mathbf{0} & \mathbf{0} & \mathbf{0} & \mathbf{0} & \mathbf{0} & \mathbf{0} & \mathbf{0} \\
        
        \mathbf{0} & \mathbf{0} & \mathbf{0} & \mathbf{0} & \mathbf{0} & \mathbf{0} & \mathbf{0} & \mathbf{0} & \mathbf{0} & \mathbf{0} & \mathbf{\delta_{mm'}\delta_{pp'}} & \mathbf{0} & \mathbf{0} & \mathbf{0} & \mathbf{0} & \mathbf{0} & \mathbf{0} & \mathbf{0} \\
    \end{bmatrix},
\end{equation*}

\noindent
where $\mathbf{\delta_{mm'}}$ and $\mathbf{\delta_{pp'}}$ are Kronecker deltas such that 

\begin{equation*}
    \mathbf{\delta}_{ij} = 
    \begin{cases} 
        1 & \text{if } i = j \\
        0 & \text{if } i \neq j.
    \end{cases}
\end{equation*}
We also define
\begin{equation*}
    \mathbf{v}_m = \underbrace{\begin{bmatrix}
    \mathbf{0}_{m-1} & 1 & \mathbf{0}_{M-m}
    \end{bmatrix}}_{P \ \text{times}}, 
\end{equation*}
and
\begin{equation*}
    \mathbf{v}_p =  \begin{bmatrix}
    \mathbf{0}_{(p-1)M}   & \mathbf{1}_M & \mathbf{0}_{(P-p)M}
    \end{bmatrix},
\end{equation*}
where $\mathbf{0}_n \in \RR^n$ and $\mathbf{1}_n \in \RR^n$ represent vectors of length $n$ filled with zeros and ones, respectively. We define $\mathbf{b}_t$ as:
\begin{equation*}
    \mathbf{b}_t = 
    \begin{bmatrix}
        \mathbf{0}_{MP} \\
        \mathbf{1}_M \\
        \mathbf{\hat{F}}_p \\
        \mathbf{0}_{MP} \\
        \mathbf{\hat{w}}_p^m(t-1)  \\
        \mathbf{Q}_{\text{sys}, MP}
    \end{bmatrix},
\end{equation*}
where $\mathbf{0}_{MP}$ is a zero vector of size $MP$, $\mathbf{1}_M$ is a vector of ones of size $M$, 

\begin{equation*}
    \mathbf{\hat{F}}_p =
    \begin{bmatrix}
        \hat{F}_1(t) \\
        \hat{F}_2(t) \\
        \vdots \\
        \hat{F}_P(t)
    \end{bmatrix}_{P \;\times\; 1},
\end{equation*}

and 
\begin{equation*}
    \mathbf{\hat{w}}_p^m(t-1) =
    \begin{bmatrix}
        \hat{w}_0^0(t-1) \\
        \hat{w}_0^1(t-1) \\
        \vdots \\
        \hat{w}_P^M(t-1)
    \end{bmatrix}_{P \;\times\; M}.
\end{equation*}

We define $\mathbf{C}$ as:
\begin{align*}
    \mathbf{C} &= 
    \begin{bmatrix}
        \mathbf{0} & \mathbf{0} & \mathbf{0} & \mathbf{0} & \mathbf{0} & \mathbf{0} & \mathbf{0} & -\mathbf{\delta_{mm'}\delta_{pp'}} & \mathbf{0} & \mathbf{0} & \mathbf{0} & \mathbf{0} & \mathbf{0} & \mathbf{0} & \mathbf{0} & \mathbf{0} & \mathbf{0} & \mathbf{0} \\
        \mathbf{0} & \mathbf{0} & \mathbf{0} & \mathbf{0} & \mathbf{0} & \mathbf{0} & \mathbf{0} & \mathbf{\delta_{mm'}\delta_{pp'}} & \mathbf{0} & \mathbf{0} & \mathbf{0} & \mathbf{0} & \mathbf{0} & \mathbf{0} & \mathbf{0} & \mathbf{0} & \mathbf{0} & \mathbf{0} \\
        \mathbf{0} & -\mathbf{\delta_{mm'}\delta_{pp'}} & \mathbf{0} & \mathbf{0} & \mathbf{0} & \mathbf{0} & \mathbf{0} & \mathbf{\delta_{mm'}\delta_{pp'}} & \mathbf{0} & \mathbf{0} & \mathbf{0} & \mathbf{0} & \mathbf{0} & \mathbf{0} & \mathbf{0} & \mathbf{0} & \mathbf{0} & \mathbf{0} \\
        \mathbf{0} & \mathbf{\delta_{mm'}\delta_{pp'}} & \mathbf{0} & \mathbf{0} & \mathbf{0} & \mathbf{0} & \mathbf{0} & -\mathbf{\delta_{mm'}\delta_{pp'}} & \mathbf{0} & \mathbf{0} & \mathbf{0} & \mathbf{0} & \mathbf{0} & \mathbf{0} & \mathbf{0} & \mathbf{0} & \mathbf{0} & \mathbf{0} \\
        \mathbf{0} & \mathbf{0} & \mathbf{0} & \mathbf{0} & \mathbf{0} & \mathbf{0} & \mathbf{0} & \frac{-1}{\Delta s}\mathbf{\delta_{mm'}\delta_{pp'}} & \mathbf{0} & \mathbf{\delta_{mm'}\delta_{pp'}} & \mathbf{0} & \mathbf{0} & \mathbf{0} & \mathbf{0} & \mathbf{0} & \mathbf{0} & \mathbf{0} & \mathbf{0} \\
        \mathbf{0} & \mathbf{0} & \mathbf{0} & \mathbf{0} & \mathbf{0} & \mathbf{0} & \mathbf{0} & \mathbf{0} & \mathbf{0} & \mathbf{0} & \mathbf{v_m} & \mathbf{0} & \mathbf{0} & \mathbf{0} & \mathbf{0} & \mathbf{0} & \mathbf{0} & \mathbf{0} \\
        \mathbf{0} & \mathbf{0} & \mathbf{0} & \mathbf{0} & \mathbf{0} & \mathbf{0} & \mathbf{0} & \mathbf{0} & \mathbf{0} & \mathbf{\delta_{mm'}\delta_{pp'}} & \mathbf{0} & \mathbf{0} & \mathbf{0} & \mathbf{0} & \mathbf{0} & \mathbf{0} & \mathbf{0} & \mathbf{0}
    \end{bmatrix}. 
\end{align*}

We define $\mathbf{d}$ as:
\begin{align*}
    \mathbf{d} &= 
    \begin{bmatrix}
        \mathbf{0} \\
        \mathbf{1} \\
        \mathbf{\overline{\Delta w}} \\
        \mathbf{\overline{\Delta w}} \\
        \mathbf{0} \\
        \mathbf{\overline{Q}} \\
        \mathbf{\overline{C}}
    \end{bmatrix}.
\end{align*}

We define $\mathbf{c}^\top$ as
\begin{equation*} %\label{eq:c} \tag{v}
\mathbf{c}^\top=
    \begin{bmatrix}
        \mathbf{0} & \mathbf{0} & \mathbf{0} & \mathbf{0} & \mathbf{0} & \mathbf{0} & \mathbf{0} & \mathbf{0} & \mathbf{v}_k & \mathbf{0} & \mathbf{0} & \mathbf{0} & \mathbf{0} & \mathbf{0} & \mathbf{v}_k & \mathbf{0} & \mathbf{0} & \mathbf{0} 
    \end{bmatrix},
\end{equation*}

such that
\begin{equation*}
    \mathbf{v}_k =  \begin{bmatrix}
    \underbrace{k_1}_{1\times M} & \underbrace{k_2}_{1\times M} & \cdots & \underbrace{k_P}_{1\times M}
    \end{bmatrix},
\end{equation*}

and where $k_p$ is the cost of losing a packet of priority $p$.

\section*{Funding Sources}
This work is supported by MDA Space, the Consortium for Research and Innovation in Aerospace in Québec (CRIAQ), the Natural Sciences and Engineering Research Council of Canada (NSERC), and the Institute for Data Valorization (IVADO).
\bibliography{sample}

\end{document}